\documentclass[journal]{IEEEtran}
\usepackage{stmaryrd}
\usepackage{amsfonts}
\usepackage{amsmath}
\usepackage{amssymb}
\usepackage{cite}
\usepackage{graphicx}
\usepackage{subfigure}
\usepackage{amsthm}
\usepackage{algorithmic}
\usepackage{algorithm}
\usepackage{stfloats}

\newtheorem{Theorem}{Theorem}
\newtheorem{Definition}{Definition}
\newtheorem{Lemma}[Theorem]{Lemma}

\def\tmu{\doteq}
\def\pmu{\equiv}

\begin{document}
%
\title{Efficient Generation of Random Bits from Finite State Markov Chains}

%
%
%

\author{Hongchao~Zhou
        and~Jehoshua~Bruck,~\IEEEmembership{Fellow,~IEEE}
\thanks{Hongchao~Zhou and Jehoshua~Bruck are with the Department
of Electrical Engineering, California Institute of Technology, Pasadena,
CA 91125, USA, e-mail: hzhou@caltech.edu; bruck@caltech.edu.}
\thanks{This work was supported in part by the NSF Expeditions in Computing
Program under grant CCF-0832824.}
}

\maketitle

\begin{abstract}

The problem of random number generation from an uncorrelated random source (of unknown probability distribution) dates back to von Neumann's 1951 work.
Elias (1972) generalized von Neumann's scheme and showed how to achieve optimal efficiency in unbiased random bits generation. Hence, a natural question is what if the sources are correlated? Both Elias and Samuelson proposed methods for generating unbiased random bits in the case of correlated sources (of unknown probability distribution), specifically, they considered finite Markov chains. However, their proposed methods are not efficient or have implementation difficulties.  Blum (1986) devised an algorithm for efficiently generating random bits from degree-2 finite Markov chains in expected linear time, however, his beautiful method is still far from optimality on information-efficiency.
In this paper, we generalize Blum's algorithm to arbitrary degree finite Markov chains and combine it with Elias's method for efficient generation of unbiased bits. As a result, we provide the first known algorithm that generates unbiased random bits from an arbitrary finite Markov chain, operates \emph{in expected linear time} and achieves the information-theoretic upper bound on efficiency.

\end{abstract}

\begin{IEEEkeywords}
Random sequence, Random bits  generation, Markov chain.
\end{IEEEkeywords}

%
\IEEEpeerreviewmaketitle

\section{Introduction}

The problem of random number generation dates back to von Neumann \cite{Neumann1951} who considered the problem of simulating an unbiased coin by using a biased coin with unknown probability. He observed that when one focuses on a pair of coin tosses, the events  $HT$ and $TH$ have the same probability ($H$ is for `head' and $T$ is for `tail'); hence, $HT$ produces the output symbol $0$ and $TH$ produces the output symbol $1$. The other two possible events, namely, $HH$ and $TT$, are ignored, namely, they do not produce any output symbols. More efficient algorithms for generating random bits from a biased coin were proposed by Hoeffding and Simons \cite{Hoeffding1970}, Elias \cite{Elias1972}, Stout and Warren \cite{Warren1984} and Peres \cite{Peres1992}.  Elias \cite{Elias1972} was the first to devise an optimal procedure in terms of the information efficiency, namely, the expected number of unbiased random bits generated per coin toss is asymptotically equal to the entropy of the biased coin. In addition, Knuth and Yao \cite{Knuth1976} presented a simple procedure for generating sequences with arbitrary probability distributions from an unbiased coin (the probability of $H$ and $T$ is $1 \over 2$). Han and Hoshi \cite{Han1997} generalized this approach and considered the case where the given coin has an arbitrary known bias.

In this paper, we study the problem of generating random bits from an arbitrary and unknown finite Markov chain (the transition matrix is unknown). The input to our problem is a sequence of symbols that represent a random trajectory through the states of the Markov chain - given this input sequence our algorithm generates an independent unbiased
binary sequence called the output sequence. This problem was first studied by Samuelson \cite{Samuelson1968}. His approach was to focus on a single state (ignoring the other states) treat the transitions out of this state as the input process, hence, reducing the problem of correlated sources to the problem of a single `independent' random source; obviously, this method is not efficient. Elias \cite{Elias1972} suggested to utilize the sequences related to all states: Producing an `independent' output sequence from the transitions out of every state and then pasting (concatenating) the collection of output sequences to generate a long output sequence. However, neither Samuelson nor Elias proved that their methods work for arbitrary Markov chains, namely, they did not prove that the transitions out of each state are independent. In fact, Blum \cite{Blum1986} probably
realized it, as he mentioned that: (i) ``Elias's algorithm is excellent, but certain difficulties arise in trying to use it (or the original von Neumann scheme) to generate bits in expected linear time from a Markov chain", and (ii) ``Elias has suggested a way to use all the symbols produced by a MC (Markov Chain). His algorithm approaches the maximum possible efficiency for a one-state MC. For a multi-state MC, his algorithm produces arbitrarily long finite sequences. He does not, however, show how to paste these finite sequences together to produce \emph{infinitely} long independent unbiased sequences."
Blum \cite{Blum1986} derived a beautiful algorithm to generate random bits from a degree-2 Markov chain \emph{in expected linear time} by utilizing the  von Neumann scheme for generating random bits from biased coin flips. While his approach can be extended to arbitrary out-degrees (the general Markov chain model used in this paper), the information-efficiency is still far from being optimal due to the low information-efficiency of the von Neumann scheme.

In this paper, we generalize Blum's algorithm to arbitrary degree finite Markov chains and combine it with existing methods for efficient generation of unbiased bits from biased coins, such as Elias's method.  As a result, we provide the first known algorithm that generates unbiased random bits from arbitrary finite Markov chains, operates in expected linear time and achieves the information-theoretic upper bound on efficiency.  Specifically, we propose an algorithm (that we call Algorithm $A$), that is a simple modification of Elias's suggestion to generate random bits, it operates on finite sequences and its efficiency can asymptotically reach the information-theoretic upper bound for long input sequences. In addition, we propose a second algorithm, called Algorithm $B$, that is a combination of Blum's and Elias's algorithms, it generates infinitely long sequences of random bits in expected linear time. One of our key ideas for generating random bits is that we explore equal-probability sequences of the same length. Hence,
a natural question is: Can we improve the efficiency by utilizing as many as possible equal-probability sequences? We provide a positive answer
to this question and describe Algorithm $C$, that is the first known polynomial-time and optimal algorithm (it is optimal in terms of information-efficiency for an arbitrary input length) for random bits generation from finite Markov chains.

In this paper, we use the following notations:
$$\begin{array}{lcl}
x_a & : & \textrm{the $a^{th}$ element of $X$}\\
X[a] & : & \textrm{same as $x_a$, the $a^{th}$ element of $X$}\\
X[a:b] &: & \textrm{subsequence of $X$ from the $a^{th}$ to $b^{th}$ element}\\
X^a & : &  X[1:a]\\
X*Y &:& \textrm{the concatenation of $X$ and $Y$}\\
& & e.g. \quad s_1s_2*s_2s_1=s_1s_2s_2s_1\\
Y \pmu X &:& Y \textrm{ is a permutation of } X \\
& & e.g. \quad s_1s_2s_2s_3\pmu s_3s_2s_2s_1\\
Y \tmu X &:& Y \textrm{ is a permutation of } X \textrm{ and } y_{|Y|}=x_{|X|}\\
& & \textrm{namely the last element is fixed}\\
& & e.g. \quad s_1s_2s_2s_3\tmu s_2s_2s_1s_3 \textrm{ where } s_3 \textrm{ is fixed}\\
\end{array}
$$

The remainder of this paper is organized as follows. Section \ref{Section_biasedcoin} reviews existing schemes for generating
random bits from arbitrarily biased coins. Section \ref{Section_MainLemma} discusses the challenge in generating random bits from arbitrary finite Markov chains and presents our main lemma - this lemma characterizes the exit sequences of Markov chains. Algorithm $A$ is presented and analyzed in Section \ref{section_finite}, it is related to Elias's ideas for generating random bits from Markov chains. Algorithm $B$ is presented in Section  \ref{section_infinite}, it is a generalization of Blum's algorithm.
An optimal algorithm, called Algorithm $C$, is described in Section \ref{section_optimal}. Finally, Section \ref{section_experiment} provides numerical evaluations of our algorithms.

\section{Generating Random Bits for Biased Coins}
\label{Section_biasedcoin}

Consider a sequence of length $N$ generated by a biased n-face coin
$$X=x_1x_2...x_N\in \{s_1,s_2,...,s_n\}^N$$
such that the probability to get $s_i$ is $p_i$, and $\sum_{i=1}^n p_i=1$.
While we are given a sequence $X$ the probabilities that $p_1,p_2,...,p_n$ are unknown, the question is: How can we efficiently generate an independent and unbiased sequence of $0$'s and $1$'s
from $X$? The efficiency (information-efficiency) of a generation algorithm is defined as the ratio between the
expected length of the output sequence and the length of the input sequence, namely, the expected number of random bits generated per input symbol. In this section we describe three existing solutions for the problem of random bits generation from biased coins.

\subsection{The von Neumann Scheme}

In 1951, von Neumann \cite{Neumann1951} considered this question for biased coins and described
a simple procedure for generating an independent unbiased binary sequence $z_1z_2...$ from the input sequence $X=x_1x_2...$. In his original procedure, the coin is binary, however, it can be simply generalized for the case of an $n$-face coin: For an input sequence, we can divide it into pairs $x_1x_2, x_3x_4, ...$ and use the following mapping for each pair
$$s_is_j(i<j)\rightarrow 0,\quad s_is_j(i>j)\rightarrow 1, \quad  s_is_i \rightarrow \phi$$
where $\phi$ denotes the empty sequence. As a result, by concatenating the outputs of all the pairs, we can get a binary sequence which is independent and unbiased.
The von Neumann scheme is computationally (very) efficient, however, its information-efficiency is far from being optimal. For example, when
the input sequence is binary, the probability for a pair of input bits to generate
an output bit (not a $\phi$) is $2p_1p_2$, hence the efficiency is $p_1p_2$, which
is $\frac{1}{4}$ at $p_1=p_2=\frac{1}{2}$ and less elsewhere.

\subsection{The Elias Scheme}

In 1972, Elias \cite{Elias1972} proposed an optimal (in terms of efficiency) algorithm
as a generalization of the von Neumann scheme; for the sake of completeness we describe it here.

Elias's method is based on the following idea: The possible $n^N$ input sequences of length $N$ can be partitioned into classes
such that all the sequences in the same class have the same number of $s_k$'s with $1\leq k\leq n$. Note that for every class, the members of the class have the same probability to be generated. For example, let $n=2$ and $N=4$, we can divide the possible $n^N=16$
input sequences into 5 classes:
\begin{eqnarray*}
  S_0&=& \{s_1s_1s_1s_1\}\\
  S_1&=& \{s_1s_1s_1s_2,s_1s_1s_2s_1,s_1s_2s_1s_1,s_2s_1s_1s_1\}\\
  S_2&=& \{s_1s_1s_2s_2,s_1s_2s_1s_2,s_1s_2s_2s_1,\\
  &&s_2s_1s_1s_2,s_2s_1s_2s_1,s_2s_2s_1s_1\}\\
  S_3&=&\{s_1s_2s_2s_2,s_2s_1s_2s_2,s_2s_2s_1s_2,s_2s_2s_2s_1\}\\
  S_4&=&\{s_2s_2s_2s_2\}
\end{eqnarray*}

Now, our goal is to assign a string of bits (the output) to each possible input sequence, such that any two output sequences $Y$ and $Y'$ with the same length (say $k$), have the same probability to be generated, namely $\frac{c_k}{2^n}$ for some $0\leq c_k\leq 1$. The idea is that for any given class we partition the members of the class to groups of sizes that are a power of 2, for a group with $2^i$ members (for some $i$) we assign binary strings of length $i$. Note that when the class size is odd we have to exclude one member of this class. We now demonstrate the idea by continuing
the example above.

Note that in the example above, we cannot assign any bits to the sequence in $S_0$, so if the input sequence is $s_1s_1s_1s_1$,
the output sequence should be $\phi$ (denotes the empty sequence). There are $4$ sequences in $S_1$ and we assign the binary strings as follows:
$$s_1s_1s_1s_2\rightarrow 00,\quad s_1s_1s_2s_1\rightarrow 01$$
$$s_1s_2s_1s_1\rightarrow 10,\quad s_2s_1s_1s_1\rightarrow 11$$
Similarly, for $S_2$, there are $6$ sequences that can be divided into a group of $4$
and a group of $2$:
$$s_1s_1s_2s_2\rightarrow 00, \quad  s_1s_2s_1s_2\rightarrow 01$$
$$s_1s_2s_2s_1\rightarrow 10, \quad  s_2s_1s_1s_2\rightarrow 11$$
$$s_2s_1s_2s_1\rightarrow 0, \quad  s_2s_2s_1s_1\rightarrow 1$$

In general, for a class with $W$ members that were not assigned yet, assign $2^j$ possible output binary sequences of length $j$ to $2^j$ distinct unassigned members, where
$2^j\leq W<2^{j+1}$.  Repeat the procedure above for the rest of the members
that were not assigned. Note that when a class has an odd number of members, there will be one and only one member assigned to $\phi$.

Given an input sequence $X$ of length $N$, using the method above, the output sequence can be written as a function of $X$, denoted by $\Psi_{E}(X)$, called the Elias function.
In \cite{Ryabko2000}, Ryabko and Matchikina showed that the Elias function of an input sequence
of length $N$ (that is generated by a biased coin with two faces) is computable in $O(N \log^3 N \log\log (N))$ time.
We can prove that their conclusion is valid in the general case of a coin with $n$ faces with $n>2$.

\subsection{The Peres Scheme}

In 1992, Peres \cite{Peres1992} demonstrated that iterating the original von Neumann scheme
on the discarded information can asymptotically achieve optimal efficiency.
Let's define the function related to the von Neumann scheme as $\Psi_1: \{0,1\}^*\rightarrow
\{0,1\}^*$. Then the iterated procedures $\Psi_v$ with $v\geq 2$ are defined inductively.
Given input sequence $x_1x_2...x_{2m}$, let $i_1<i_2<...<i_k$ denote all the indices $i\leq m$ for which $x_{2i}=x_{2i-1}$, then $\Psi_v$ is defined as
\begin{eqnarray*}
 && \Psi_v(x_1,x_2,...,x_{2m}) \\
  &=& \Psi_1(x_1,x_2,...,x_{2m})*\Psi_{v-1}(x_1\oplus x_2, ..., x_{2m-1}\oplus x_{2m}) \\
   && *\Psi_{v-1}(x_{i_1}, ..., x_{i_k})
\end{eqnarray*}

Note that on the righthand side of the equation above, the first term corresponds to the random bits
generated with the von Neumann scheme, the second and third terms relate
to the symmetric information discarded by the von Neumann scheme.

Finally, we can define $\Psi_v$ for sequences of odd length by
$$\Psi_v(x_1,x_2,...,x_{2m+1})=\Psi_v(x_1,x_2,...,x_{2m})$$

Surprisingly, this simple iterative procedure achieves the optimal efficiency asymptotically.
The computational complexity and memory requirements of this scheme are substantially smaller than those of the Elias scheme. However, a drawback of this scheme is that its generalization to the case of an $n$-face coin with $n>2$ is not obvious.

\subsection{Properties of the Schemes}

Let's denote $\Psi: \{s_1,s_2,...,s_n\}^N \rightarrow \{0,1\}^*$ as a scheme that generates independent unbiased sequences from any biased coins (with unknown probabilities).
Such $\Psi$ can be the von Neumann scheme, the Elias scheme, the Peres scheme or any other scheme. Let $X$ be a sequence generated from an arbitrary biased coin, with length $N$, then
a property of $\Psi$ is that for any $Y\in \{0,1\}^*$ and $Y' \in \{0,1\}^*$ with $|Y|=|Y'|$, we have
$$P[\Psi(X)=Y]=P[\Psi(X)=Y']$$
Namely, two output sequences of equal length have equal probability.


That leads to the following property for $\Psi$. It says that given the number of $s_i$'s for all $i$ with $1\leq i\leq n$,
the number of such sequences to yield a binary sequence $Y$ equals to that of sequences to yield $Y'$ if
$Y$ and $Y'$ have the same length. It further implies that given the condition of knowing the number of $s_i$'s for all $i$ with $1\leq i\leq n$, the output sequence
of $\Psi$ is still independent and unbiased. This property is due to the
linear independence of probability functions of the sequences with different numbers
of the $s_i$'s.

\begin{Lemma}
Let $S$ be a subset in $\{s_1,s_2, ..., s_n\}^N$ such that it includes all the sequences with the same number of $s_i$'s for all $i$ with $1\leq i\leq n$, namely, $k_1, k_2, ..., k_n$. Let $B_Y$ denote the set $\{X|\Psi(X)=Y\}$. Then for any $Y\in \{0,1\}^*$ and $Y' \in \{0,1\}^*$ with $|Y|=|Y'|$, we have $|S\bigcap B_Y|=|S\bigcap B_{Y'}|$. \label{lemma_coin}
\end{Lemma}

\proof In $S$, the number of $s_i$'s in each sequence is $k_i$ for all $1\leq i\leq n$, then we can get that
$$P[\Psi(X)=Y]=\sum_S |S\bigcap B_Y|\prod_{i=1}^n \beta(S) $$
where
$$\beta(S)=\prod_{i=1}^n p_i^{k_i}$$

Since $P[\Psi(X)=Y]=P[\Psi(X)=Y']$, we have
$$\sum_S(|S\bigcap B_Y|-|S\bigcap B_{Y'}|)\beta(S)=0$$

The set of polynomials $\bigcup_{S}\{\beta(S)\}$ is linearly independent in the vector space of functions on $[0,1]$, so we can conclude that
$|S\bigcap B_Y|=|S\bigcap B_{Y'}|$.
\hfill\QED

\begin{table*}
  \centering
  \begin{tabular}{|c|l|c|c|}
  \hline
\textrm{Input sequence}\quad & \textrm{Probability}\quad & $\Psi(\pi_1(X))$ & $\Psi(\pi_1(X))*\Psi(\pi_2(X))$\\
\hline
  $s_1s_1s_1s_1$& $(1-p_1)^3$ & $\phi$& $\phi$\\
  $s_1s_1s_1s_2$& $(1-p_1)^2 p_1$ & $0$ & $0$\\
  $s_1s_1s_2s_1$& $(1-p_1)p_1p_2$ & $0$& $0$\\
  $s_1s_1s_2s_2$& $(1-p_1)p_1(1-p_2)$ & $0$& $0$\\
  $s_1s_2s_1s_1$& $p_1p_2(1-p_1)$ & $1$& $1$\\
  $s_1s_2s_1s_2$& $p_1^2p_2$ & $\phi$& $\phi$\\
  $s_1s_2s_2s_1$& $p_1(1-p_2)p_2$ & $\phi$ & 1\\
  $s_1s_2s_2s_2$& $p_1(1-p_2)^2$ & $\phi$ & $\phi$\\
  \hline
\end{tabular}
\caption{Probabilities of exit sequences - an example that simple concatenation does not work.}\label{table11}
\end{table*}

\section{Some Properties of Markov Chains}
\label{Section_MainLemma}

Our goal is to efficiently generate random bits from a Markov chain with unknown transition probabilities.
The paradigm we study is that a Markov chain generates the sequence of states that it is visiting and
this sequence of states is the input sequence to our algorithm for generating random bits.
Specifically, we express an input sequence as $X=x_1x_2...x_N$ with $x_i\in \{s_1,s_2,...,s_n\}$, where
$\{s_1,s_2,...,s_n\}$ indicate the states of a Markov chain.

One idea is that for a given Markov chain, we can treat each state, say $s$, as a coin and consider the `next states' (the states the chain has transitioned to after being at state $s$) as the results of a coin toss. Namely, we can generate a collection of sequences $\pi(X)=[\pi_1(X),\pi_2(X),...,\pi_n(X)]$, called exit sequences, where $\pi_i(X)$ is the sequence of states following $s_i$ in $X$, namely,
$$\pi_i(X) =\{x_{j+1}|x_j=s_i, 1\leq j<N\}$$
For example, assume that the input sequence is
$$X=s_1 s_4 s_2 s_1 s_3 s_2 s_3 s_1 s_1 s_2 s_3 s_4 s_1$$
If we consider the states following $s_1$ we get $\pi_1(X)$ as the set of states in boldface:
$$X = s_1 \textbf{s}_\textbf{{4}} s_2 s_1 \textbf{s}_\textbf{{3}} s_2 s_3 s_1 \textbf{s}_\textbf{{1}} \textbf{s}_\textbf{{2}} s_3 s_4 s_1$$
Hence, the exit sequences are:
\begin{eqnarray*}
\pi_1(X)&=&s_4 s_3 s_1s_2\\
\pi_2(X)&=&s_1 s_3 s_3\\
\pi_3(X)&=&s_2s_1 s_4\\
\pi_4(X)&=&s_2 s_1
\end{eqnarray*}

\begin{Lemma}[Uniqueness]
An input sequence $X$ can be uniquely determined by $x_1$ and $\pi(X)$.\label{lemma_Uniqueness}
\end{Lemma}
\proof
Given $x_1$ and $\pi(X)$, according to the work of Blum in \cite{Blum1986}, $x_1x_2...x_N$ can uniquely
be constructed in the following way: Initially, set the starting state as $x_1$. Inductively, if $x_i=s_k$,
then set $x_{i+1}$ as the first element in $\pi_k(X)$ and remove the first element of $\pi_k(X)$. Finally, we can uniquely generate the sequence $x_1x_2...x_N$.
\hfill\QED

\begin{Lemma}[Equal-probability]  Two input sequences $X=x_1x_2...x_N$ and $Y=y_1y_2...y_N$ with $x_1=y_1$ have the same probability to be generated if $\pi_i(X)\pmu \pi_i(Y)$ for all $1\leq i\leq n$.\label{lemma_equalP}
\end{Lemma}
\proof
Note that the probability to generate $X$ is
$$P[X]=P[x_1]P[x_2|x_1]...P[x_N|x_{N-1}]$$
and the probability to generate $Y$ is
$$P[Y]=P[y_1]P[y_2|y_1]...P[y_N|y_{N-1}]$$
By permutating the terms in the expression above, it is not hard to get that $P[X]=P[Y]$
if $x_1=y_1$  and $\pi_i(X)\pmu \pi_i(Y)$ for all $1\leq i\leq n$. Basically, the exit sequences
describe the edges that are used in the trajectory in the Markov chain. The edges in the trajectories
that correspond to $X$ and $Y$ are identical, hence $P[X]=P[Y]$.
\hfill\QED

\begin{figure}[!h]
\centering
\includegraphics[width=2.5in]{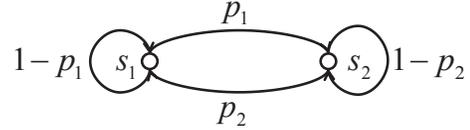}
\caption{An example of Markov chain with two states.}
\label{fig_MCexample2}
\end{figure}

In \cite{Samuelson1968}, Samuelson considered a two-state Markov chain, and he pointed out that it may generate unbiased random bits by applying the von Neumann scheme to the exit sequence of state $s_1$. Later, in \cite{Elias1972}, in order to increase the efficiency, Elias has suggested a scheme that uses all the symbols produced by a Markov chain. His main idea was to create the final output sequence by concatenating
the output sequences that correspond to $\pi_1(X),\pi_2(X),...$. However, neither Samuelson nor Elias proved that their methods produce random output sequences that are independent and unbiased, in fact, their proposed methods are not correct for some cases.
To demonstrate it we consider: (1)  $\Psi(\pi_1(X))$ as the final output. (2)  $\Psi(\pi_1(X))*\Psi(\pi_2(X))*...$ as the final output.
For example, consider the two-state Markov chain in which $P[s_2|s_1]=p_1$ and $P[s_1|s_2]=p_2$, as shown in Fig. \ref{fig_MCexample2}.

Assume that an input sequence of length $N=4$ is generated from this Markov chain and the starting state is $s_1$, then the probabilities of the possible input sequences and their corresponding output sequences are given in Table \ref{table11}. In the table we can see that the probabilities to produce $0$ or $1$ are different for some $p_1$ and $p_2$ in both methods, presented in columns 3 and 4, respectively.

The problem of generating random bits from an arbitrary Markov chain is challenging, as Blum said in \cite{Blum1986}: ``Elias's algorithm is excellent, but certain difficulties arise in trying to use it (or the original von Neumann scheme) to generate random bits in expected linear time from a Markov chain". It seems that the exit sequence of a state is independent since each exit of the state will not affect the other exits. However, this is not always true
when the length of the input sequence is given, say $N$. Let's still consider the example of
a two-state Markov chain in Fig. \ref{fig_MCexample2}. Assume the starting state of this Markov chain is $s_1$, if $1-p_1>0$, then with non-zero probability we have
$$\pi_1(X)=s_1s_1...s_1$$
whose length is $N-1$. But it is impossible to have
$$\pi_1(X)=s_2s_2...s_2$$
of length $N-1$. That means $\pi_1(X)$ is not an independent sequence. The main reason is that
although each exit of a state will not affect the other exits, it will affect the length of the exit sequence.
In fact, $\pi_1(X)$ is an independent sequence if the length of $\pi_1(X)$ is given, instead of giving the length of $X$.

In this paper, we consider this problem from another perspective. According to Lemma \ref{lemma_equalP}, we know that permutating the exit sequences does not change the probability of a sequence, however, the permuted sequence has to correspond to a trajectory in the Markov chain. The reason for this contingency is that in some cases the permuted sequence does not correspond to a trajectory: Consider the following example,
$$X=s_1 s_4 s_2 s_1 s_3 s_2 s_3 s_1 s_1 s_2 s_3 s_4 s_1$$
and
$$\pi(X)=[s_4 s_3 s_1s_2,s_1 s_3 s_3,s_2s_1 s_4, s_2 s_1]$$
If we permutate the last exit sequence $s_2s_1$ to $s_1s_2$, we cannot get a new sequence such that
its starting state is $s_1$ and its exit sequences are
$$[s_4 s_3 s_1s_2,s_1 s_3 s_3,s_2s_1 s_4, s_1 s_2]$$
This can be verified by attempting to construct the sequence using
Blum's method (which is given in the proof of Lemma \ref{lemma_Uniqueness}).
Notice that if we permutate the first exit sequence $s_4s_3s_1s_2$ into $s_1s_2s_3s_4$, we can find such a new sequence, which is
$$Y=s_1s_1s_2s_1s_3s_2s_3s_1s_4s_2s_3s_4s_1$$
This observation motivated us to study
the characterization of exit sequences that are feasible in Markov chains (or finite state machines).

\begin{Definition}[Feasibility]
Given a Markov chain, a starting state $s_\alpha$  and a collection of sequences $\Lambda=[\Lambda_1,\Lambda_2,...,\Lambda_n]$,
we say that $(s_\alpha,\Lambda)$ is feasible if and only if there exists a sequence $X$ that corresponds to a trajectory in the Markov chain such that $x_{1}= s_\alpha$ and $\pi(X)=\Lambda$.
\end{Definition}

Based on the definition of feasibility, we present the main technical lemma of the paper. Repeating the notation from the beginning of the paper, we say that a sequence $Y$ is a tail-fixed permutation of $X$, denoted as $Y\tmu X$, if
and only if (1) $Y$ is a permutation of $X$, and (2) $X$ and $Y$ have the same last element, namely, $y_{|Y|}=x_{|X|}$.

\begin{Lemma}[Main Lemma: Feasibility and equivalence of exit sequences]
Given a starting state $s_\alpha$ and two collections of sequences $\Lambda=[\Lambda_1,\Lambda_2,...,\Lambda_n]$ and $\Gamma=[\Gamma_1,\Gamma_2,...,\Gamma_n]$
such that $\Lambda_i \tmu \Gamma_i$ (tail-fixed permutation) for all $1\leq i\leq n$. Then $(s_\alpha, \Lambda)$ is feasible if and only if $(s_\alpha, \Gamma)$ is feasible.
\label{lemma_main1}
\end{Lemma}

The proof of this main lemma will be given in the Appendix.  According to the main lemma, we have the following equivalent statement.

\begin{Lemma}[Feasible permutations of exit sequences]
Given an input sequence $X=x_1x_2...x_N$ with $x_N=s_\chi$ that produced from a Markov chain. Assume that $[\Lambda_1,\Lambda_2,...,\Lambda_n]$
is an aribitrary collection of exit sequences that corresponds to the exit sequences of $X$ as follows:
\begin{enumerate}
  \item $\Lambda_i$ is a permutation ($\pmu$) of $\pi_i(X)$,  for  $i=\chi$.
  \item $\Lambda_i$ is a tail-fixed permutation  ($\tmu$) of  $\pi_i(X)$, for  $i\neq \chi$.
\end{enumerate}
Then there exists a feasible sequence $X'=x_1'x_2'...x_N'$ such that
$x_1'=x_1$ and $\pi(X')=[\Lambda_1,\Lambda_2,...,\Lambda_n]$. For this $X'$, we have $x_N'=x_N$.\label{lemma_main2}
\end{Lemma}

One might reason that Lemma \ref{lemma_main2} is stronger than the main lemma (Lemma \ref{lemma_main1}). However, we will show that these
two lemmas are equivalent. It is obvious that if the statement in Lemma \ref{lemma_main2} is true, then the main lemma is also true. Now we show that
if the main lemma is true then the statement in Lemma \ref{lemma_main2} is also true.

\proof Given  $X=x_1x_2...x_N$, let's add one more symbol $s_{n+1}$ to the end of $X$ ($s_{n+1}$ is different from all the states in $X$), then we can get a new sequence $x_1x_2...x_Ns_{n+1}$, whose exit sequences are
$$[\pi_1(X),\pi_2(X),...,\pi_{\chi}(X)s_{n+1},...,\pi_n(X),\phi]$$

According to the main lemma, we know that there exists another sequence $x_1'x_2'...x_N'x_{N+1}'$ such that its exit sequences are
$$[\Lambda_1,\Lambda_2,...,\Lambda_{\chi}s_{n+1},...\Lambda_n, \phi]$$ and $x_1'=x_1$. Definitely, the last symbol of this sequence is $s_{n+1}$, i.e.,
$x_{N+1}'=s_{n+1}$. As a result, we have $x_N'=s_\chi$.

Now, by removing the last element from $x_1'x_2'...x_N'x_{N+1}'$, we can get a new sequence $x=x_1'x_2'...x_N'$ such that its exit
sequences are $$[\Lambda_1,\Lambda_2,...,\Lambda_{\chi},...\Lambda_n]$$
and  $x_1'=x_1$. We also have $x_N'=s_\chi$.

This completes the proof.
\hfill\QED

We demonstrate the result above by considering
the example at the beginning of this section. Let
$$X=s_1s_4s_2 s_1 s_3s_2 s_3 s_1 s_1s_2s_3s_4s_1$$
with $\chi=1$ and its exit sequences is given by
$$[s_4 s_3 s_1s_2, s_1s_3s_3, s_2s_1s_4, s_2s_1]$$
After permutating all the exit sequences (for $i\neq 1$, we keep the last element of the $i^{th}$ sequence fixed),
we get a new group of exit sequences
$$[s_1s_2s_3s_4 , s_3s_1s_3, s_1s_2s_4, s_2s_1]$$
Based on these new exit sequences, we can generate a new input sequence
$$X'=s_1s_1s_2s_3s_1s_3s_2s_1s_4s_2s_3s_4s_1$$
This accords with the statements above.

\section{Algorithm A : Modification of Elias's Suggestion}
\label{section_finite}

In the section above, we see that Elias suggested to paste the outputs of different exit sequences together, as the final output, but
the simple direct concatenation cannot always work. By modifying the method of pasting these outputs, we get Algorithm $A$
to generate unbiased random bits from any Markov chains.

\begin{list}{\labelitemi}{\leftmargin=0.5em}
\renewcommand{\labelitemi}{}
  \item
  \item \textbf{Algorithm A}
  \item \textbf{Input:} A sequence $X=x_1x_2...x_N$ produced by a Markov chain, where $x_i\in S=\{s_1,s_2,...,s_n\}$.
  \item \textbf{Output:} A sequence of $0'$s and $1'$s.
  \item \textbf{Main Function:}
  \begin{algorithmic}
\STATE Suppose $x_N=s_\chi$.
\FOR{$i:=1 \textrm{ to } n$}
    \IF{$i=\chi$}
        \STATE Output $\Psi(\pi_i(X))$.
    \ELSE
        \STATE Output $\Psi(\pi_i(X)^{|\pi_i(X)|-1})$
    \ENDIF
\ENDFOR
\end{algorithmic}
\item \textbf{Comment:} (1) $\Psi(X)$ can be any scheme that generates random bits from biased coins. For example, we can use the Elias function.
(2) When $i=\chi$, we can also output $\Psi(\pi_i(X)^{|\pi_i(X)|-1})$ for simplicity, but the efficiency may be reduced a little.
\item
\end{list}

The only difference between Algorithm $A$  and direct concatenation is that: Algorithm $A$ ignores the last symbols of some exit sequences. Let's go back to the example of a two-state Markov chain with $P[s_2|s_1]=p_1$ and $P[s_1|s_2]=p_2$ in Fig. \ref{fig_MCexample2}, which demonstrates that
direct concatenation does not always work well. Here, we still assume that an input sequence with length $N=4$ is generated from this Markov chain and the starting state is $s_1$, then
the probability of each possible input sequence and its corresponding output sequence (based on Algorithm $A$) are given by:

\vspace{0.5cm}
\begin{tabular}{|c|l|c|}
  \hline
\textrm{Input sequence}\quad & \textrm{Probability}\quad & \textrm{Output sequence}\\
\hline
  $s_1s_1s_1s_1$& $(1-p_1)^3$ & $\phi$\\
  $s_1s_1s_1s_2$& $(1-p_1)^2 p_1$ & $\phi$\\
  $s_1s_1s_2s_1$& $(1-p_1)p_1p_2$ & $0$\\
  $s_1s_1s_2s_2$& $(1-p_1)p_1(1-p_2)$ & $\phi$\\
  $s_1s_2s_1s_1$& $p_1p_2(1-p_1)$ & $1$\\
  $s_1s_2s_1s_2$& $p_1^2p_2$ & $\phi$\\
  $s_1s_2s_2s_1$& $p_1(1-p_2)p_2$ & $\phi$\\
  $s_1s_2s_2s_2$& $p_1(1-p_2)^2$ & $\phi$\\
  \hline
\end{tabular}
\vspace{0.5cm}

We can see that when the input sequence length $N=4$, a bit $0$ and a bit $1$ have the same probability to be generated and no longer sequences are generated.
In this case, the output sequence is independent and unbiased.

In order to prove that  all the sequences generated by Algorithm $A$ are independent and unbiased, we need to show that for any sequences $Y$ and $Y'$ of the same length, they have the same probability to be generated.

\begin{Theorem}[Algorithm A]
Let the sequence
generated by a Markov chain be used as input to Algorithm $A$, then the
output of Algorithm $A$ is an independent unbiased sequence.\label{theorem_AlgorithmA}
\end{Theorem}

\proof Let's first divide all the possible sequences in $\{s_1,s_2,...,s_n\}^N$ into groups, and use $G$ to denote the set of the groups. Two sequences $X$ and $X'$
are in the same group if and only if
\begin{enumerate}
  \item $x_1'=x_1$ and $x_N'=x_N=s_\chi$ for some $\chi$.
  \item If $i=\chi$, $\pi_i(X')\pmu \pi_i(X)$.
  \item If $i\neq \chi$, $\pi_i(X')\tmu \pi_i(X)$.
\end{enumerate}
We will show that for each group $S\in G$, the number of sequences to generate $Y$ equals to that of sequences to generate $Y'$ if $Y$ and $Y'$ have the same length, i.e., $|S\bigcap B_Y|=|S\bigcap B_{Y'}|$
if $|Y|=|Y'|$, where $B_Y$ is the set of sequences of length $N$ that yield $Y$.

Now, given a group $S$, if $i=\chi$ let's define $S_i$ as the set of
all the permutations of $\pi_i(X)$ for $X\in S$, and if $i\neq \chi$ let's define $S_i$ as the set of all the permutations of $\pi_i(X)^{|\pi_i(X)|-1}$ for $X\in S$. According to Lemma \ref{lemma_coin}, we know that for any $Y,Y'\in \{0,1\}^{l}$,
there are  the same number of members in $S_i$ which generate $Y$ and $Y'$. So we can use
$|S_i(l)|$ to denote the number of members in $S_i$ which generate a certain binary sequence with length $l$ (e.g. $Y$).

According to the definitions above, let $l_1,l_2,...,l_n$ be non-negative integers, then we have
$$|S\bigcap B_Y|=\sum_{l_1+...+l_n=|Y|}\prod_{i=1}^n|S_i(l_i)|$$
where each combination $(l_1,l_2,...,l_n)$ is a partition of the length of $Y$.

Similarly, we also have
$$|S\bigcap B_{Y'}|=\sum_{l_1+...+l_n=|Y'|}\prod_{i=1}^n|S_i(l_i)|$$
which tells us that $|S\bigcap B_Y|=|S\bigcap B_{Y'}|$ if $|Y|=|Y'|$.

Note that all the sequences in the same group $S$ have the same probability to be generated. So when $|Y|=|Y'|$, the
probability to generate $Y$ is
\begin{eqnarray*}
&& P[X\in B_Y]\\
&=&\sum_{S\in G} P[S] \sum_{X\in S}P[X\in B_Y|X\in S]\\
&=&\sum_{S\in G} P[S] \sum_{X\in S}\frac{|S\bigcap B_Y|}{|S|}\\
&=&\sum_{S\in G} P[S] \sum_{X\in S}\frac{|S\bigcap B_{Y'}|}{|S|}\\
&=& P[X\in B_{Y'}]
\end{eqnarray*}
which implies that output sequence is independent and unbiased.
\hfill\QED

\begin{Theorem}[Efficiency]
Let $X$ be a sequence of length $N$ generated by a Markov chain, which is used as input to Algorithm $A$. Let $\Psi$ in Algorithm $A$ be Elias's function.
Suppose the length of its output sequence is $M$, then the limiting efficiency $
\eta_N=\frac{E[M]}{N}$ as ${N\rightarrow\infty}$ realizes the upper bound $\frac{H(X)}{N}$.
\label{theorem_efficiencyA}
\end{Theorem}

\proof Here, the upper bound $\frac{H(X)}{N}$ is provided by Elias \cite{Elias1972}. We can use the same argument in Elias's paper \cite{Elias1972} to prove this theorem.

Let $X_i$ denote the next state of $s_i$. Obviously, $X_i$ is a random variable for $1\leq i\leq n$, whose
entropy is denoted as $H(X_i)$.  Let $U= (u_1,u_2, \ldots, u_n)$ denote the stationary distribution of the Markov chain, then
we have
$$\lim_{N\rightarrow \infty}\frac{H(X)}{N}=\sum_{i=1}^n u_i H(X_i)$$

When $N\rightarrow \infty$, there exists an $\epsilon_N$ which $\rightarrow 0$, such that
with probability $1-\epsilon_N$, $|\pi_i(X)|>(u_i-\epsilon_N)N$ for all $1\leq i\leq n$. Using Algorithm A, with probability $1-\epsilon_N$, the length $M$ of the output sequence is bounded below by
$$\sum_{i=1}^n (1-\epsilon_N)(|\pi_i(X)|-1)\eta_i$$
where $\eta_i$ is the efficiency of the $\Psi$ when the input is $\pi_i(X)$ or $\pi_i(X)^{|\pi_i(X)|-1}$.
According to Theorem 2 in Elias's paper \cite{Elias1972}, we know that as $|\pi_i(X)|\rightarrow \infty$, $\eta_i\rightarrow H(X_i)$.
So with probability $1-\epsilon_N$, the length $M$ of the output sequence  is below bounded by
$$\sum_{i=1}^N (1-\epsilon_N)((u_i-\epsilon_N)N-1)(1-\epsilon_N) H(X_i)$$

Then we have
\begin{eqnarray*}
&&\lim_{N\rightarrow\infty}\frac{E[M]}{N}\\
&\geq&\lim_{N\rightarrow\infty} \frac{[\sum_{i=1}^N (1-\epsilon_N)^3((u_i-\epsilon_N)N-1)H(X_i)]}{N}\\
&=& \lim_{N\rightarrow \infty }\frac{H(X)}{N}
\end{eqnarray*}
At the same time, $\frac{E[M]}{N}$ is upper bounded by $\frac{H(X)}{N}$. So we can get
$$\lim_{N\rightarrow\infty}\frac{E[M]}{N}=\lim_{N\rightarrow \infty }\frac{H(X)}{N}$$
which completes the proof. \hfill\QED

Given an input sequence, it is efficient to generate
independent unbiased sequences using Algorithm $A$. However, it has some limitations: (1) The complete input sequence has to be stored.
(2) For a long input sequence it is computationally intensive as it depends on the input length.
(3) The method works for finite-length sequences and does not lend itself to stream processing.
In order to address these limitations we propose two variants of Algorithm $A$.

In the first variant of Algorithm $A$, instead of applying $\Psi$ directly to $\Lambda_i=\pi_i(X)$ for $i=\chi$ (or $\Lambda_i=\pi_i(X)^{|\pi_i(X)|-1}$ for $i\neq\chi$), we first
split $\Lambda_i$ into several segments with lengths $k_{i1}, k_{i2}, ...$ then apply $\Psi$ to all of the segments separately. It can be proved that this variant of Algorithm A
can generate independent unbiased sequences from an arbitrary Markov chain, as long as $k_{i1}, k_{i2}, ...$ do not depend on the order of elements in each exit sequence.
For example, we can split $\Lambda_i$ into two segments of lengths $\lfloor\frac{|\Lambda_i|}{2}\rfloor$ and $\lceil\frac{|\Lambda_i|}{2}\rceil$, we can also split
it into three segments of lengths $(a,a,|\Lambda_i|-2a)$ ... Generally, the shorter each segment is, the faster we can obtain the final output. But at the same time, we may have to
sacrifice a little information efficiency.

The second variant of Algorithm $A$ is based on the following idea: for a given sequence from a Markov chain, we can
split it into some shorter sequences such that they are independent of each other, therefore we can apply Algorithm $A$ to all of the sequences and then
concatenate their output sequences together as the final one. In order to do this, given a sequence $X=x_1x_2...$, we can use $x_1=s_\alpha$
as a special state to it. For example, in practice, we can set a constant $k$, if there exists a minimal integer $i$ such that $x_i=s_\alpha$ and $i>k$, then
we can split $X$ into two sequences $x_1x_2...x_i$ and $x_ix_{i+1}...$ (note that both of the sequences have the element $x_i$). For the second sequence $x_ix_{i+1}...$,
we can repeat the some procedure ... Iteratively, we can split a sequence $X$ into several sequences such that they are independent of each other. These sequences, with the exception of the last one, start and end with $s_\alpha$, and their lengths are usually slightly longer than $k$.

\section{Algorithm B : Generalization of Blum's Algorithm}
\label{section_infinite}

In \cite{Blum1986}, Blum proposed a beautiful algorithm to generate an independent unbiased sequence of $0$'s and $1$'s from any Markov chain by extending von Neumann scheme.
His algorithm can deal with infinitely long sequences and use only constant space and expected linear time. The only drawback of his algorithm is that its efficiency is still far from
the information-theoretic upper bound, due to the limitation (compared to the
Elias algorithm) of the von Neumann scheme. In this section, we generalize Blum's algorithm by replacing von Neumann scheme with
Elias's. As a result, we get Algorithm $B$: It maintains some good properties of Blum's algorithm and its efficiency approaches the information-theoretic upper bound.

\begin{list}{\labelitemi}{\leftmargin=0.5em}
\renewcommand{\labelitemi}{}
  \item
  \item \textbf{Algorithm B}
  \item \textbf{Input:} A sequence (or a stream) $x_1x_2...$ produced by a Markov chain, where $x_i\in \{s_1,s_2,...,s_n\}$.
  \item \textbf{Parameter:} $n$ positive integer functions (window size) $\varpi_i(k)$ with $k\geq 1$ for $1\leq i\leq n$.
  \item \textbf{Output:} A sequence (or a stream) of $0$'s and $1$'s.
  \item \textbf{Main Function:}
\begin{algorithmic}
\STATE $E_i=\phi$ (empty) for all $1\leq i\leq n$.
\STATE $k_i=1$ for all $1\leq i\leq n$.
\STATE $c:$ the index of current state, namely, $s_c=x_1$.
\WHILE{next input symbol is $s_j$ ($\neq $ \textbf{null})}
\STATE $E_c=E_cs_j$ (Add $s_j$ to $E_{c}$).
\IF{$|E_j|\geq \varpi_j(k_j)$}
\STATE Output $\Psi(E_j)$.
\STATE $E_j=\phi$.
\STATE $k_j=k_j+1$.
\ENDIF
\STATE $c=j$.
\ENDWHILE
\end{algorithmic}
\item
\end{list}

In the algorithm above, we apply function $\Psi$ on $E_j$ to generate
random bits if and only if the window for $E_j$ is completely filled and
the Markov chain is currently at state $s_j$.

For example, we set $\varpi_i(k)=4$ for all $1\leq i\leq n$ and the input sequence is $$X=s_1s_1s_1s_2s_2s_2s_1s_2s_2$$

After reading the last second ($8^{th}$) symbol $s_2$, we have
$$E_1=s_1s_1s_2s_2 \quad E_2=s_2s_2s_1$$
In this case, $|E_1|\geq 4$ so the window for $E_1$ is full, but we don't apply $\Psi$ to $E_1$ because
the current state of the Markov chain is $s_2$, not $s_1$.

By reading the last ($9^{th}$) symbol $s_2$, we get
$$E_1=s_1s_1s_2s_2 \quad E_2=s_2s_2s_1s_2$$
Since the current state of the Markov chain is $s_2$ and $|E_2|\geq 4$,
we produce $\Psi(E_2=s_2s_2s_1s_2)$ and reset $E_2$ as $\phi$.

In the example above, treating $X$ as input to Algorithm $B$, we can get the output sequence is
$\Psi(s_2s_2s_1s_2)$. The algorithm does not output $\Psi(E_1=s_1s_1s_2s_2)$ until the Markov chain reaches state $s_1$ again. Timing is crucial!

Note that Blum's algorithm is a special case of Algorithm $B$ by setting the window size functions $\varpi_i(k)=2$ for all $1\leq i\leq n$ and $k\in \{1,2,...\}$.
Namely, Algorithm $B$ is a generalization of Blum's algorithm, the key is that when we increase the windows sizes, we can apply more efficient schemes (compared to the von Neumann scheme) for $\Psi$.
Assume a sequence of symbols $X=x_1x_2...x_N$ with $x_N=s_\chi$ have been read by the algorithm above, we want to show that for any $N$, the output
sequence is always independent and unbiased. Unfortunately, Blum's
proof for the case of $\varpi_i(k)=2$ cannot be applied to our proposed scheme.

\begin{figure}[!h]
\centering
\includegraphics[width=2.8in]{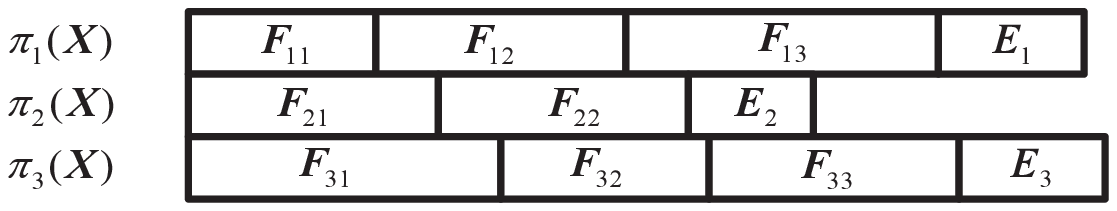}
\caption{The simplified expressions  for the exit sequences of $X$.}
\label{fig_example2}
\end{figure}

For all $i$ with $1\leq i\leq n$, we can write
$$\pi_i(X)=F_{i1}F_{i2}...F_{im_i}E_i$$
where $F_{ij}$ with $1\leq j\leq m_i$ are the segments used to generate outputs.
For all $i,j$, we have
$$|F_{ij}|=\varpi_i(j)$$
and
$$\left\{\begin{array}{cc}
           0\leq |E_i|< \varpi_i(m_i+1) & \textrm{ if } i=\chi \\
           0< |E_i|\leq \varpi_i(m_i+1) & \textrm{ otherwise }
         \end{array}
\right.$$
See Fig. \ref{fig_example2} for simple illustration.

\begin{Theorem}[Algorithm $B$]
Let the sequence
generated by a Markov chain be used as input to Algorithm $B$, then Algorithm $B$
generates an independent unbiased sequence of bits in expected linear time.
\end{Theorem}

\proof In the following proof, we use the same idea as in the proof for Algorithm $A$.

Let's first divide all the possible input sequences in $\{s_1, s_2, ..., s_n\}^N$ into groups, and use $G$ to denote the group set. Two sequences $X$ and $X'$ are in the same group if and only if

\begin{enumerate}
\item $x_1=x'_1$ and $x_N=x'_N$.
  \item For all $i$ with $1\leq i\leq n$,
  $$\pi_i(X)=F_{i1}F_{i2}...F_{im_i}E_i$$
$$\pi_i(X')=F'_{i1}F'_{i2}...F'_{im_i}E'_i$$
where $F_{ij}$ and $F'_{ij}$ are the segments used to generate outputs.
  \item For all $i,j$, $F_{ij}\pmu F'_{ij}$.
  \item For all $i$, $E_i=E'_i$.
\end{enumerate}

We will show that in each group $S\in G$, the number of sequences to generate $Y$ equals to that of sequences to generate $Y'$ if $|Y|=|Y'|$, i.e., $|S\bigcap B_Y|=|S\bigcap B_{Y'}|$
for $|Y|=|Y'|$, where $B_Y$ is the set of sequences of length $N$ that yield $Y$.

Now, given a group $S$, let's define $S_{ij}$ as the set of
all the permutations of $F_{ij}$ for $X\in S$. According to Lemma \ref{lemma_coin}, we know that for different $Y\in \{0,1\}^{l}$,
there are  the same number of members in $S_{ij}$ which generate $Y$. So we can use
$|S_{ij}(l)|$ to denote the number of members in $S_{ij}$ which generate a certain binary sequence with length $l$.

Let $l_{11},l_{12},...,l_{1m_1},l_{21}...,l_{nm_n}$ be non-negative integers such that their sum is $|Y|$,
we want to prove that
$$|S\bigcap B_Y|=\sum_{l_{11}+...+l_{nm_n}=|Y|}\prod_{i=1}^n\prod_{j=1}^{m_i}|S_{ij}(l_{ij})|$$
The proof is by induction. Let $w=\sum_{i=1}^n m_i$. First, the conclusion holds
for $w=1$. Assume the conclusion holds for $w>1$, we want to prove that
the conclusion also holds for $w+1$.

Given an $X\in S$, assume $F_{im_i}$ is the last segment that generates an output. According to our main lemma (Lemma \ref{lemma_main1}), we know that for any sequence in $S$, $F_{im_i}$ is always the last segment that generates an output.
Now, let's fix $F_{im_i}$ and assume $F_{im_i}$ generates the last $l_{im_i}$ bits of $Y$. We want
to know how many sequences in $S\bigcap B_Y$ have $F_{im_i}$ as their last segments that generate outputs?
In order to get the answer, we concatenate $F_{im_i}$ with $E_i$ as the new $E_i$. As a result,
we have $\sum_{i=1}^n m_i-1=w$ segments to generate the first $|Y|-l_{im_i}$ bits of $Y$. Based on our assumption, the number of such sequences will be
$$\sum_{l_{11}+...+l_{i(m_i-1)}+...=|Y|-l_{im_i}}\frac{1}{|S_{im_i}(l_{im_i})|}\prod_{k=1}^n\prod_{j=1}^{m_i}|S_{kj}(l_{kj})|$$
where $l_{11}, ..., l_{i(m_i-1)}, l_{(i+1)1} ..., l_{nm_n}$ are non-negative integers.
For a different $l_{im_i}$, there are $|S_{im_i}(l_{im_i})|$ choices for $F_{im_i}$. Therefore,
$|S\bigcap B_Y|$ can be obtained by multiplying $|S_{im_i}(l_{im_i})|$ by the number above and summing
them up over $l_{im_i}$. Namely, we can get the conclusion above.

According to this conclusion, we know that if $|Y|=|Y'|$, then $|S\bigcap B_Y|=|S\bigcap B_{Y'}|$. Using the same argument as in
Theorem \ref{theorem_AlgorithmA} we complete the proof of the theorem.
\hfill\QED

Normally, the window size functions $\varpi_i(k)$ for $1\leq i\leq n$ can be any positive integer functions. Here,
we fix these window size functions as a constant, namely, $\varpi$. By increasing the value of $\varpi$, we can increase the efficiency of the scheme, but
at the same time it may cost more storage space and need more waiting time. It is helpful to
analyze the relationship between scheme efficiency and window size $\varpi$.

\begin{Theorem}[Efficiency]
Let $X$ be a sequence of length
$N$ generated by a Markov chain with transition matrix $P$, which is used as input
to Algorithm $B$ with constant window size $\varpi$. Then as the length of the sequence goes to infinity,
the limiting efficiency of Algorithm $B$ is
$$\eta(\varpi)=\sum_{i=1}^{n} u_i\eta_{i}(\varpi)$$
where $U=(u_1,u_2,...,u_n)$ is the stationary distribution of this Markov chain, and
$\eta_{i}(\varpi)$ is the efficiency of $\Psi$ when the input sequence of length $\varpi$ is generated
by a $n$-face coin with distribution $(p_{i1},p_{i2},...,p_{in})$.
\end{Theorem}

\proof  When $N\rightarrow \infty$, there exists an $\epsilon_N$ which $\rightarrow 0$, such that
with probability $1-\epsilon_N$, $(u_i-\epsilon_N)N<|\pi_i(X)|<(u_i+\epsilon_N)N$ for all $1\leq i\leq n$.

The efficiency of Algorithm $B$ can be written as $\eta(\varpi)$, which satisfies
$$\frac{\sum_{i=1}^{n} \lfloor\frac{|\pi_i(X)|-1}{\varpi}\rfloor \eta_i(\varpi)\varpi}{N}\leq \eta(\varpi)\leq \frac{\sum_{i=1}^{n} \lfloor\frac{|\pi_i(X)|}{\varpi}\rfloor \eta_i(\varpi)\varpi}{N}$$

With probability $1-\epsilon_N$, we have
{\small $$\frac{\sum_{i=1}^{n} (\frac{(u_i-\epsilon_N)N}{\varpi}-1) \eta_i(\varpi)\varpi}{N}\leq \eta(\varpi) \leq \frac{\sum_{i=1}^{n}\frac{ (u_i-\epsilon_N)N}{\varpi} \eta_i(\varpi)\varpi}{N}$$}

So when $N\rightarrow \infty$, we have that
$$\eta(\varpi)=\sum_{i=1}^{n} u_i\eta_{i}(\varpi)$$

This completes the proof.
\hfill\QED

Let's define $\alpha(N)=\sum n_k 2^{n_k}$, where $\sum 2^{n_k}$ is the standard binary expansion of $N$. Assume $\Psi$ is the Elias function, then
$$\eta_i(\varpi)=\frac{1}{\varpi}\sum_{k_1+...+k_n=\varpi}\alpha(\frac{\varpi!}{k_1!k_2!...k_n!})p_{i1}^{k_1}p_{i2}^{k_2}...p_{in}^{k_n}$$
Based on this formula, we can numerically study the relationship between the limiting efficiency and the window size  (see Section \ref{section_experiment}). In fact, when the window size becomes large, the limiting efficiency ($n\rightarrow\infty$) approaches the information-theoretic upper bound.

\section{Algorithm C : An Optimal Algorithm}
\label{section_optimal}

Both Algorithm $A$ and Algorithm $B$ are asymptotically optimal, but when the length of the input sequence is finite they may not be optimal.
In this section, we try to construct an optimal algorithm, called Algorithm C, such that its information-efficiency is maximized when the length of the input sequence is finite. Before presenting this algorithm, following the idea of Pae and Loui\cite{Pae2005}, we first discuss the equivalent condition for a function $f$ to generate random bits from an arbitrary Markov chain, and then
present the sufficient condition for $f$ to be optimal.

\begin{Lemma}[Equivalent condition]\label{lemma_equ_condition}
Let $K=\{k_{ij}\}$ be an $n\times n$ non-negative integer matrix with $\sum_{i=1}^n\sum_{j=1}^n k_{ij}=N-1$. We define $S_{(\alpha,K)}$ as
$$S_{(\alpha,K)}=\{X\in\{s_1,s_2,...,s_n\}^N| k_j(\pi_i(X))=k_{ij}, x_1=s_{\alpha}\}$$
where $k_j(X)$ is the number of $s_j$ in $X$.
A function  $f:\{s_1,s_2,...,s_n\}^N \rightarrow \{0,1\}^*$ can generate random bits from an arbitrary Markov chain, if and only if for any $(\alpha,K)$ and two
binary sequences $Y$ and $Y'$ with $|Y|=|Y'|$,
$$|S_{(\alpha, K)}\bigcap B_Y|=|S_{(\alpha, K)}\bigcap B_{Y'}|$$
where $B_Y=\{X|X\in \{s_1,s_2,...,s_n\}^N, f(X)=Y\}$ is the set of sequences of length $N$ that yield $Y$.
\end{Lemma}

\proof If $f$ can generate random bits from an arbitrary Markov chain, then $P[f(X)=Y]=P[f(X)=Y']$ for any two binary sequences $Y$ and $Y'$ of the same length.
Here, we can write
$$P[f(X)=Y]=\sum_{\alpha, K}|S_{(\alpha,K)}\bigcap B_Y|\phi(K)P(x_1=s_\alpha)$$
where
$\phi(K)=\prod_{i=1}^n \prod_{j=1}^n p_{ij}^{k_{ij}}$
and $\phi(K)P(x_1=s_\alpha)$ is
the probability to generate a sequence with starting state $s_{\alpha}$ and with exit sequences specified by $K$ if such input sequence exists.
Similarly,
$$P[f(X)=Y']=\sum_{\alpha, K}|S_{(\alpha,K)}\bigcap B_{Y'}|\phi(K)P(x_1=s_\alpha)$$
As a result, $$\sum_{(\alpha,K)}(|S_{(\alpha,K)}\bigcap B_{Y'}|-|S_{(\alpha,K)}\bigcap B_Y|)\phi(K)P(x_1=s_\alpha)=0$$

Since $P(x_1=s_\alpha)$ can be any value in $[0,1]$, for all $1\leq \alpha \leq n$ we have
$$\sum_{K}(|S_{(\alpha,K)}\bigcap B_{Y'}|-|S_{(\alpha,K)}\bigcap B_Y|)\phi(K)=0$$

According to the linear independence of $\bigcup_{K}\{\phi(K)\}$ in the vector space of functions on $[0,1]$,
we can conclude that $|S_{(\alpha,K)}\bigcap B_Y|=|S_{(\alpha,K)}\bigcap B_{Y'}|$ for all $(\alpha,K)$ if $|Y|=|Y'|$.

Inversely, if for all $Y,Y'$ with the same length, $|S_{(\alpha,K)}\bigcap B_Y|=|S_{(\alpha,K)}\bigcap B_{Y'}|$ for all $(\alpha,K)$, then $Y$ and $Y'$ have the same probability to be generated.
Therefore, $f$ can generate random bits from an arbitrary Markov chain.
\hfill\QED

Let's define $\alpha(N)=\sum n_k 2^{n_k}$, where $\sum 2^{n_k}$ is the standard binary expansion of $N$, then we have the sufficient condition for an optimal function .

\begin{Lemma}[Sufficient condition for an optimal function]
Let $f^*$ be a function that generates random bits from an arbitrary Markov chain with unknown transition probabilities. If for any $\alpha$ and any $n\times n$ non-negative integer matrix $K$ with $\sum_{i=1}^n\sum_{j=1}^n k_{ij}=N-1$, the following equation is satisfied,
$$\sum_{X\in S_{(\alpha,K)}}|f^*(X)|=\alpha(|S_{(\alpha,K)}|)$$
then $f^*$ generates independent unbiased random bits with optimal information efficiency. Note that $|f^*(X)|$ is the length of $f^*(x)$ and $|S_{(\alpha,K)}|$ is the size of $S_{(\alpha,K)}$. \label{lemma_conditionForOptimal}
\end{Lemma}

\proof
Let $h$ denote an arbitrary function that is able to generate random bits from any Markov chain. According to Lemma 2.9 in \cite{Pae2005}, we know that
$$ \sum_{X\in S_{(\alpha,K)}}|h(X)| \leq \alpha(|S_{(\alpha,K)}|)$$

Then the average output length of $h$ is
\begin{eqnarray*}
   E(|h(X)|)&=& \frac{1}{N}\sum_{(\alpha,K)}\sum_{X\in S_{(\alpha,K)}}|h(X)|\phi(K)P[x_1=s_\alpha] \\
   &\leq & \frac{1}{N}\sum_{(\alpha,K)} \alpha(|S_{(\alpha,K)}|) \phi(K)P[x_1=s_\alpha] \\
   &=& \frac{1}{N}\sum_{(\alpha,K)} \sum_{X\in S_{(\alpha,K)}}|f^*(X)| \phi(K)P[x_1=s_\alpha]\\
   &=& E(|f^*(X)|)
\end{eqnarray*}
So $f^*$ is the optimal one. This completes the proof.
\hfill\QED

Here, we construct the following algorithm (Algorithm $C$) which satisfies all the conditions in Lemma \ref{lemma_equ_condition} and Lemma \ref{lemma_conditionForOptimal}.
As a result, it can generate unbiased random bits from an arbitrary Markov chain with optimal information efficiency.

\begin{list}{\labelitemi}{\leftmargin=0.5em}
\renewcommand{\labelitemi}{}
  \item
  \item \textbf{Algorithm C}
  \item \textbf{Input:} A sequence $X=x_1x_2...,x_N$ produced by a Markov chain, where $x_i\in S=\{s_1,s_2,...,s_n\}$.
  \item \textbf{Output:} A sequence of $0'$s and $1'$s.
  \item \textbf{Main Function:}
\begin{algorithmic}
\STATE \textbf{1)} Get the matrix $K=\{k_{ij}\}$ with
$$k_{ij}=k_j(\pi_i(X))$$
\STATE \textbf{2)} Define $S(X)$ as
$$S(X)=\{X'| k_j(\pi_i(X'))=k_{ij} \forall i,j; x_1'=x_1\}$$
then compute $|S(X)|$.
\STATE \textbf{3)} Compute the rank $r(X)$ of $X$ in $S(X)$ with respect to a given order.
\STATE \textbf{4)} According to $|S(X)|$ and $r(X)$, determine the output sequence.
Let $\sum_{k}2^{n_k}$ be the standard binary expansion of $|S(X)|$ with $n_1>n_2>...$
and assume the
starting value of $r(X)$ is $0$. If $r(X)<2^{n_1}$, the output is the $n_1$ digit
binary representation of $r(x)$. If $\sum_{k=1}^i 2^{n_k}\leq r(x)< \sum_{k=1}^{i+1} 2^{n_k}$, the output is the $n_{i+1}$ digit
binary representation of $r(x)$.
\end{algorithmic}
\item \textbf{Comment:} The fast calculations of $|S(X)|$ and $r(x)$ will be given in the rest of this section.
\end{list}

In Algorithm $A$, when we use Elias's function as $\Psi$, the limiting efficiency $
\eta_N=\frac{E[M]}{N}$ (as $N\rightarrow\infty$) realizes the bound $\frac{H(X)}{N}$. Algorithm $C$ is optimal, so it has the same or higher efficiency.
Therefore, the limiting efficiency of Algorithm $C$ as $N\rightarrow\infty$ also realizes the bound $\frac{H(X)}{N}$.

In Algorithm $C$, for an input sequence $X$ with $x_N=s_\chi$, we can rank it with respect to the lexicographic order of $\theta(X)$ and $\sigma(X)$.
Here, we define
$$\theta(X)=(\pi_1(X)_{|\pi_1(X)|},\ldots,\pi_{n}(X)_{|\pi_{n}(X)|})$$
which is the vector of the last symbols of $\pi_i(X)$ for $1\leq i\leq n$. And
$\sigma(X)$ is the complement of $\theta(X)$ in $\pi(X)$, namely,
$$\sigma(X)=(\pi_1(X)^{|\pi_1(X)|-1},\ldots,\pi_{n}(X)^{|\pi_{n}(X)|-1})$$

For example, when the input sequence is
$$X=s_1s_4s_2s_1s_3s_2s_3s_1s_1s_2s_3s_4s_1$$
Its exit sequences is
$$\pi(X)=[s_4s_3s_1s_2,s_1s_3s_3,s_2s_1s_4,s_2s_1]$$
Then for this input sequence $X$, we have that
$$\theta(X)=[s_2,s_3,s_4,s_1]$$
$$\sigma(X)=[s_4s_2s_1,s_1s_3,s_2s_1,s_2]$$

Based on the lexicographic order defined above, both $|S(X)|$ and $r(X)$ can be obtained using a brute-force search. However, this approach in not computationally efficient.
Here, we describe an efficient algorithm for computing $|S(X)|$ and $r(X)$, such that Algorithm $C$ is computable in $O(N \log^3 N \log\log (N))$ time. This method is inspired by the algorithm for computing the Elias function that is described in \cite{Ryabko2000}.

\begin{Lemma}
$|S(X)|$ in Algorithm $C$ is computable in $O(N(\log N\log\log N)^2)$ time.
\end{Lemma}

\proof
The idea to compute $|S(X)|$ in Algorithm $C$ is that we can divide $S(X)$ into different classes,
denoted by $S(X,\theta)$ for different $\theta$ such that
$$S(X,\theta)=\{X'|\forall i,j, k_j(\pi_i(X'))=k_{ij}, \theta(X')=\theta\}$$
where $k_{ij}=k_j(\pi_i(X))$ is the number of $s_j$'s in $\pi_i(X)$ for all
$1\leq i,j\leq n$. $\theta(X)$ is the vector of the last symbols of $\pi(X)$ defined above.
As a result, we have $|S(X)|=\sum_{\theta}|S(X,\theta)|$. Although it is not easy to calculate $|S(X)|$ directly, but it is much easier to
compute $|S(X,\theta)|$ for a given $\theta$.

For a given $\theta=(\theta_1,\theta_2,...,\theta_n)$, we need first determine whether $S(X,\theta)$ is empty or not. In order to do this, we quickly construct a collection of
exit sequences $\Lambda=[\Lambda_1,\Lambda_2,...,\Lambda_n]$ by moving the first $\theta_i$ in $\pi_i(X)$ to the end for all $1\leq i\leq n$. According to the main lemma,
we know that $S(X,\theta)$ is empty if and only if $\pi_i(X)$ does not include $\theta_i$ for some $i$ or $(x_1,\Lambda)$ is not feasible.

If $S(X,\theta)$ is not empty, then $(x_1,\Lambda)$ is feasible. In this case, based on the main lemma, we have
$$|S(X,\theta)|=\prod_{i=1}^n \frac{(k_{i1}+k_{i2}+...+k_{in}-1)!}{k_{i1}!...(k_{i\theta_i}-1)!...k_{in}!}$$
$$= (\prod_{i=1}^n \frac{(k_{i1}+k_{i2}+...+k_{in})!}{k_{i1}!k_{i2}!...k_{in}!})(\prod_{i=1}^n\frac{k_{i\theta_i}}{(k_{i1}+k_{i2}+...+k_{in})})$$
where the first term, denoted by $Z$, is computable in $O(N(\log N\log\log N)^2)$ time \cite{Borwein1985}. Further more, we can get that
$$|S(X)|=\sum_{\theta}|S(X,\theta)|=Z(\sum_{\theta}\prod_{i=1}^n\frac{k_{i\theta_i}}{(k_{i1}+k_{i2}+...+k_{in})})$$ is also computable in $O(N(\log N\log\log N)^2)$ time.
\hfill\QED

\begin{Lemma} $r(X)$ in Algorithm $C$ is computable in $O(N\log^3 N\log\log N)$ time.
\end{Lemma}

\proof
Based on some calculations in the lemma above, we can try to obtain $r(X)$ when $X$ is ranked
with respect to the lexicographic order of $\theta(X)$ and $\sigma(X)$.
Let $r(X,\theta(X))$ denote the rank of $X$ in $S(X,\theta(X))$, then we have that
$$r(X)=\sum_{\theta<\theta(X)}|S(X,\theta)|+r(X,\theta(X))$$
where $<$ is based on the lexicographic order. In the formula, $\sum_{\theta<\theta(X)}|S(X,\theta)|$ can be efficiently obtained by computing
$$Z\frac{\sum_{\theta<\theta(X):|S(X,\theta)|>0}\prod_{i=1}^n k_{i\theta_i}}{\prod_{i=1}^n(k_{i1}+k_{i2}+...+k_{in})}$$
where $Z$ is defined in the last lemma.
So far, we only need to compute $r(X,\theta(X))$, with respect to the lexicography order of $\sigma(X)$. $\sigma(X)$ can be written as a group of sequences $[\sigma_1(X),\sigma_2(X),...,\sigma_n(X)]$ such that for all $1\leq i\leq n$
$$\sigma_i(X)=\pi_i(X)^{|\pi_i(X)|-1}$$

There are $M=(N-1)-n$ symbols in $\sigma(X)$. Let $r_i(X)$ be the number of sequences $X'\in S(X, \theta(X))$ such that the first $M-i$
symbols of $\sigma(X')$ are the same with that of $\sigma(X)$ and the $M-i+1^{th}$ symbol of $\sigma(X')$ is smaller than that of $\sigma(X)$, then we can get that
$$r(X,\theta(X))=\sum_{i=1}^{M} r_i(X)$$

Assume the $M-i+1^{th}$ symbol in $\sigma(X)$ is the $u_i^{th}$ symbol in $\sigma_{v_i}(X)$. Then we can get that
$$r_i(X)=\sum_{s_w<{\sigma_{v_i}}[u_i]}\frac{k_w(T_i)}{|T_i|}\frac{|T_i|!}{k_1(T_i)!...k_n(T_i)!} \prod_{j>v_i} N_j(X)$$
where $T_i$ is the subsequence of $\sigma_{v_i}(X)$ from the $u_i^{th}$ symbol to the end; $N_j(X)$
is the number of permutations for $\sigma_j(X)$.

Let's define the values
$$\rho_i^0=\frac{|T_i|}{k_{w_i}(T_i)},\quad \lambda_i^0=\sum_{s_w<\sigma_{v_i}[u_i]} \frac{k_w(T_i)}{|T_i|}$$
where $w_i$ is the index of the first symbol of $T_i$, i.e., ${\sigma_{v_i}}[u_i]=s_{w_i}$. Then $r(X,\theta(X))$ can be written as
$$r(X,\theta(X))=\sum_{i=1}^{M} \lambda_i^0\rho_i^0\rho_{i-1}^0...\rho_1^0$$

Suppose that $\log_2 M $ is an integer. Otherwise, we can add trivial terms to the formula above to make $\log N$ an integer. In order to quickly calculate $r(X,\theta(X))$, the following calculations are performed:
$$\rho_i^s = \rho_{2i-1}^{s-1}\rho_{2i}^{s-1},\lambda_i^s=\lambda_{2i-1}^{s-1}+\lambda_{2i}^{s-1}\rho_{2i}^{s-1}$$
$$s=1,2,...,\log M; i=1,2,..., 2^{-s}M$$

Then applying the method in \cite{Ryabko2000}, we have that
$$r(X,\theta(X))=\lambda_1^{\log_2 M}$$
which is computable in $O(M\log^3M \log\log M)$ time. As a result, for a fixed $n$, $r(X)$ is computable in $O(N\log^3 N \log\log N)$ time.
\hfill\QED

Based on the discussion above, we know that Algorithm $C$ is computable in $O(N\log^3 N\log\log N)$ time.

\section{Numerical Results}
\label{section_experiment}

In this section, we describe numerical results related to the implementations of Algorithm $A$, Algorithm $B$, and Algorithm $C$. We use the Elias function for $\Psi$.

In the first experiment, we use the following randomly generated a transition matrix for a Markov chain with three states.
$$P=\left(
  \begin{array}{ccc}
0.300987 & 0.468876  & 0.230135\\
0.462996 &  0.480767 & 0.056236\\
0.42424 &  0.032404 & 0.543355
  \end{array}
\right)$$
Consider a sequence of length $12$ that is generated by the Markov chain defined above and assume that $s_1$ is the first state of this sequence. Namely, there are $3^{11}=177147$ possible input sequences. For each possible input sequence, we can compute its generating
probability and the corresponding output sequences using our three algorithms. Table \ref{Table1} presents the results of calculating the probabilities of all possible output sequences for the three algorithms. Note that the results show that indeed the outputs of the algorithms are independent unbiased sequences. Also, Algorithm $C$ has the highest information efficiency (it is optimal), and Algorithm $A$ has a higher information efficiency than Algorithm $B$ (with window size 4).

\begin{table}
  \centering
  \begin{tabular}{llll}
  \hline
  Output &  Probability  & Probability  & Probability  \\
  & Algorithm $A$ & Algorithm $B$ & Algorithm $C$ \\
  &&with $\varpi=4$&\\
  \hline
  $\Lambda$ & 0.0224191  & 0.1094849& 0.0208336\\
  0 &0.0260692   & 0.0215901& 0.0200917\\
  1 & 0.0260692  & 0.0215901&0.0200917 \\
  00 & 0.0298179 & 0.1011625&0.0206147\\
  10 & 0.0298179& 0.1011625& 0.0206147\\
  01 & 0.0298179 &0.1011625& 0.0206147\\
  11 & 0.0298179 & 0.1011625&0.0206147\\
  000 & 0.0244406 & 0.0242258& 0.0171941\\
  100 & 0.0244406& 0.0242258& 0.0171941\\
  $\ldots$ & $\ldots$& $\ldots$ & \ldots\\
011111 &0.0018831  &1.39E-5& 0.0029596\\
111111 & 0.0018831 &1.39E-5& 0.0029596\\
0000000 & 1.305E-4& & 6.056E-4\\
1000000 & 1.305E-4& &6.056E-4\\
$\ldots$ & && $\ldots$ \\
0111111 & 1.305E-4 && 6.056E-4\\
1111111 & 1.305E-4 && 6.056E-4\\
00000000 & & &1.44E-5\\
10000000 & &&1.44E-5 \\
$\ldots$ & && $\ldots$ \\
01111111&  &&  1.44E-5\\
11111111&  & &  1.44E-5\\
  \hline
Expected Length& 3.829  & 2.494& 4.355\\
\hline
\end{tabular}
  \caption{The probability of each possible output sequence and the expected output length.
  }\label{Table1}
\end{table}

\begin{figure}[!h]
\centering
\includegraphics[width=3.6in]{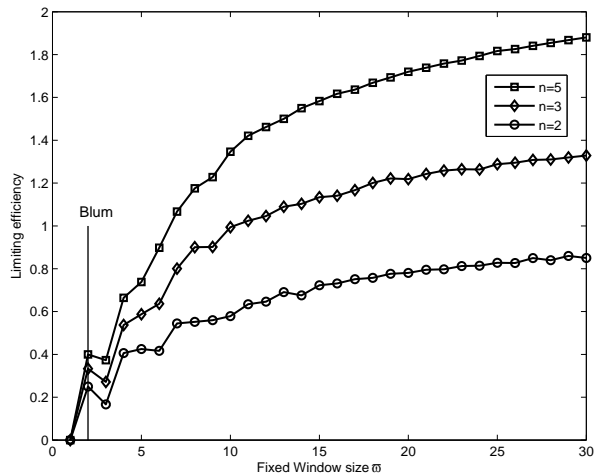}
\caption{The limiting efficiency of Algorithm $B$ varies with the value of window size $\varpi$ for different state number $n$, where we assume
that the transition probability $p_{ij}=\frac{1}{n}$ for all $1\leq i,j\leq n$.}
\label{fig_effiency}
\end{figure}

In the second calculation, we want to test the influence of window size $\varpi$ (assume $\varpi_i(k)=\varpi$ for $1\leq i\leq n$) on the efficiency of Algorithm $B$. Since the efficiency depends on the transition matrix of the Markov chain we decided to evaluate of the efficiency related to the uniform transition matrix, namely all the entries are  $\frac{1}{n}$, where $n$ is the number of states. We assume that $n$ is infinitely large. In this case, the stationary distribution of
the Markov chain is $\{\frac{1}{n},\frac{1}{n},...,\frac{1}{n}\}$.  Fig. \ref{fig_effiency} shows that when $\varpi=2$ (Blum's Algorithm), the limiting efficiencies for $n=(2,3,5)$ are $(\frac{1}{4},\frac{1}{3},\frac{2}{5})$, respectively. When $\varpi=15$, their corresponding efficiencies are
$(0.7228, 1.1342, 1.5827)$. So if the input sequence is long enough, by changing $\varpi$ from $2$ to $15$, the efficiency can increase
$189\%$ for $n=2$, $240\%$ for $n=3$ and $296\%$ for $n=4$. When $\varpi$ is small, we can increase the efficiency of Algorithm $B$ significantly  by increasing the window size $\varpi$.
When $\varpi$ becomes larger, the efficiency of Algorithm $B$ will converge to the information-theoretical upper bound, namely, $\log_2 n$. Note that $3$ is not a good value for the window size in the algorithm. That is because the Elias function is not very efficient
when the length of the input sequence is $3$. Let's consider a biased coin with two states $s_1, s_2$. If the
input sequence is $s_1s_1s_1$ or $s_2s_2s_2$, the Elias function will generate nothing. For all other
cases, it has only $2/3$ chance to generate one bit and $1/3$ chance to generate nothing. As a result,
the efficiency is even worse than the efficiency when the length of the input sequence equals to $2$.

\section{Concluding Remarks}

We considered the classical problem of generating independent unbiased bits from an arbitrary Markov chain with unknown transition probabilities. Our main contribution is the first known algorithm that has expected linear time complexity and achieves the information-theoretic upper bound on efficiency.

Our work is related to a number of interesting results in both computer science and information theory. In computer science, the attention has focused on extracting randomness from a general weak source (introduced by Zuckerman \cite{Zuc90}). Hence, the concept of an extractor was introduced - it converts weak random sequences to `random-looking' sequences, using an additional small number of truly random bits. During the past two decades, extractors and their applications have been studied extensively, see \cite{Nis96}\cite{Sha02} for surveys on the topic.  While our algorithms generate truly random bits (given a prefect Markov chain as a source) the goal of extractors is to generate `random-looking' sequences which are asymptotically close to random bits.

In information theory, it was discovered that optimal source codes can be used as universal random bits generators from arbitrary stationary ergodic random sources \cite{Visweswariah98}\cite{Han05}. When the input sequence is generated from a stationary ergodic process and it is long enough one can obtain an output sequence that behaves like truly random bits in the sense of normalized divergence. However, in some cases, the definition of normalized divergence is not strong enough. For example, suppose $Y$ is a sequence of unbiased random bits in the sense of
normalized divergence, and $1*Y$ is $Y$ with a $1$ concatenated at the beginning. If the sequence $Y$ is long enough the sequence $1*Y$
is a sequence of unbiased random bits in the sense of normalized divergence. However the sequence $1*Y$ might not
be useful in applications that are sensitive to the randomness of the first bit.

\appendix
In this appendix, we prove the main lemma.

{\hspace{-0.3cm}\textbf{Lemma 4} (Main Lemma: Feasibility and equivalence of exit sequences).
\emph
{Given a starting state $s_\alpha$ and two collections of sequences $\Lambda=[\Lambda_1,\Lambda_2,...,\Lambda_n]$ and $\Gamma=[\Gamma_1,\Gamma_2,...,\Gamma_n]$
such that $\Lambda_i \tmu \Gamma_i$ (tail-fixed permutation) for all $1\leq i\leq n$. Then $(s_\alpha, \Lambda)$ is feasible if and only if $(s_\alpha, \Gamma)$ is feasible.}
}
\vspace{0.2cm}

\begin{figure}[!h]
\centering
\includegraphics[width=2.5in]{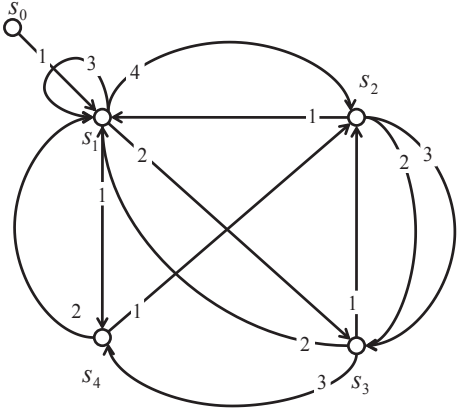}
\caption{An example of a sequence graph $G$.}
\label{fig_orderTransform}
\end{figure}

In the rest of the appendix we will prove the main lemma. To illustrate the claim in the lemma, we
express $s_\alpha$ and $\Lambda$ by a directed graph that has labels on the vertices and edges, we call this graph a \emph{sequence graph}. For example, when $s_\alpha=s_1$ and  $\Lambda=[s_4s_3s_1s_2, s_1s_3s_3,s_2s_1s_4,s_2s_1]$, we have the directed graph in Fig. \ref{fig_orderTransform}.

Let $V$ denote the vertex set, then 
$$V=\{s_0,s_1,s_2,...,s_n\}$$
and the edge set is 
$$E=\{(s_i,\Lambda_i[k])\} \bigcup \{(s_0, s_\alpha)\}$$
For each edge $(s_i,\Lambda_i[k])$, the label of this edge is $k$. For the edge $(s_0,s_\alpha)$, the label is $1$.  Namely, the label set of the outgoing edges of each state is $\{1,2,...\}$.

Given the labeling of the directed graph as defined above, we say that it contains a \emph{complete walk} if there is a path in the graph that visits all the edges, without visiting an edge twice, in the following way: (1) Start from $s_0$. (2) At each vertex, we choose an unvisited edge with the minimal label to follow. Obviously, the labeling corresponding to $(s_\alpha, \Lambda)$ is a \emph{complete walk} if and only if $(s_\alpha, \Lambda)$ is feasible. In this case, for short, we also say that $(s_\alpha, \Lambda)$ is a complete walk.
Before continuing to prove the main lemma, we first give Lemma \ref{lemma_graphTransform} and Lemma \ref{lemma_graphTransform2}.

\begin{Lemma} Assume $(s_\alpha,\Lambda)$ with $\Lambda=[\Lambda_1,\Lambda_2,...,\Lambda_\chi,...,\Lambda_n]$ is a a complete walk, which ends at state $s_\chi$. Then $(s_\alpha,\Gamma)$ with $\Gamma=[\Lambda_1,...,\Gamma_\chi,...,\Lambda_n]$ is also a complete walk ending at $s_\chi$, if
$\Lambda_\chi \pmu \Gamma_\chi$ (permutation). \label{lemma_graphTransform}
\end{Lemma}
\proof $(s_\alpha,\Lambda)$ and $(s_\alpha,\Gamma)$ correspond to different labelings on the same directed graph $G$, denoted by $L_1$ and $L_2$. Since $L_1$ is a complete walk, it can
travel all the edges in $G$ one by one, denoted as
$$(s_{i_1},s_{j_1}),(s_{i_2},s_{j_2}),...,(s_{i_N},s_{j_N})$$
where $s_{i_1}=s_0$ and $s_{j_N}=s_\chi$. We call $\{1,2,...,N\}$ as the indexes of the edges.

Based on $L_2$, let's have a walk on $G$ starting from $s_0$ until there is no unvisited outgoing edges to select.
In this walk, assume the following edges have been visited:
$$(s_{i_{w_1}},s_{j_{w_1}}),(s_{i_{w_2}},s_{j_{w_2}}),...,(s_{i_{w_M}},s_{j_{w_M}})$$
where $w_1,w_2,...,w_N$ are distinct indexes chosen from $\{1,2,...,N\}$ and $s_{i_{w_1}}=s_0$.
In order to prove that $L_2$ is a complete walk, we need to show that
(1) $s_{j_{w_M}}=s_\chi$ and (2) $M=N$.

First, let's prove that $s_{j_{w_M}}=s_\chi$. In $G$, let $N_i^{(out)}$ denote the number of outgoing edges of $s_i$ and let
$N_{i}^{(in)}$ denote the number of incoming edges of $s_i$, then we have that
$$\left\{\begin{array}{c}
           N_{0}^{(in)}=0, N_{0}^{(out)}=1 \\
           N_{\chi}^{(in)}=N_{\chi}^{(out)}+1 \\
           N_{i}^{(in)}=N_{i}^{(out)} \textrm{ for } i\neq 0, i\neq \chi
         \end{array}
\right. $$
Based on these relations, we know that once we have a walk starting from $s_0$ in $G$, this walk will finally end at state $s_\chi$.
That is because we can always get out of $s_i$ due to $N_{i}^{(in)}=N_{i}^{(out)}$ if $i\neq \chi,0$.

\begin{figure}[!h]
\centering
\includegraphics[width=2.8in]{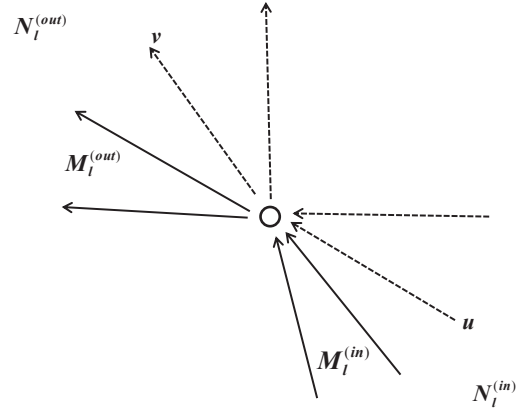}
\caption{An illustration of the incoming and outgoing edges of $s_l$. In which, the solid arrows indicate visited edges, and the dashed arrows indicate unvisited edges.}
\label{fig_example3}
\end{figure}

Now, we prove that $M=N$. This can be proved by contradiction. Assume $M\neq N$, then we define
$$V=\{w_1,w_2,...,w_M\}$$
$$\overline{V}=\{1,2,...,N\}/\{w_1,w_2,...,w_M\}$$
where $V$ corresponds to the visited edges based on $L_2$ and $\overline{V}$ corresponds to the unvisited edges based on $L_2$.
Let $v=\min (\overline{V})$, then $(s_{i_v},s_{j_v})$ is the unvisited edge with the minimal index.
Let $l=i_v$, then $(s_{i_v},s_{j_v})$ is an outgoing edge of $s_{l}$. Here $l\neq\chi$, because all the
outgoing edges of $s_\chi$ have been visited. Assume the number of visited incoming edges of $s_l$ is
$M_l^{(in)}$ and the number of visited outgoing edges of $s_l$ is $M_l^{(out)}$, then
$$M_l^{(in)}=M_l^{(out)}$$
see Fig. \ref{fig_example3} as an example.

Note that the labels of the outgoing edges of $s_l$ are the same for $L_1$ and $L_2$, since $l\neq \chi,0$. Therefore, based on $L_1$, before visiting edge $(s_{i_v},s_{j_v})$,
there must be $M_l^{(out)}$ outgoing edges of $s_l$ have been visited. As a result, based on $L_1$, there must be $M_l^{(out)}+1=M_l^{(in)}+1$ incoming edges of $s_l$ have been visited before visiting $(s_{i_v},s_{j_v})$. Among all these $M_l^{(in)}+1$
incoming edges, there exists at least one edge $(s_{i_u}, s_{j_u})$ such that $u \in \overline{V}$, since only $M_l^{(in)}$ incoming edges of $s_l$ have been visited based on $L_2$.

According to our assumption, both $u,v\in \overline{V}$ and $v$ is the minimal one, so $u>v$. On the other hand, we know that $(s_{i_u}, s_{j_u})$ is visited before $(s_{i_v},s_{j_v})$ based on $L_1$, so $u<v$. Here, the contradiction happens. Therefore, $M=N$.

This completes the proof.
\hfill\QED

Here, let's give an example of the lemma above. We know that, when $s_\alpha=s_1, \Lambda=[s_4 s_3 s_1s_2,s_1 s_3 s_3,s_2s_1 s_4, s_2 s_1]$,
$(s_\alpha, \Lambda)$ is feasible. The labeling on a directed graph corresponding to $(s_\alpha, \Lambda)$ is given in Fig. \ref{fig_orderTransform}, which is a complete walk starting at state $s_0$ and ending at state $s_1$. The path of the walk is
$$s_0s_1s_4s_2 s_1 s_3s_2 s_3 s_1 s_1s_2s_3 s_4s_1$$

\begin{figure}[!h]
\centering
\includegraphics[width=2.5in]{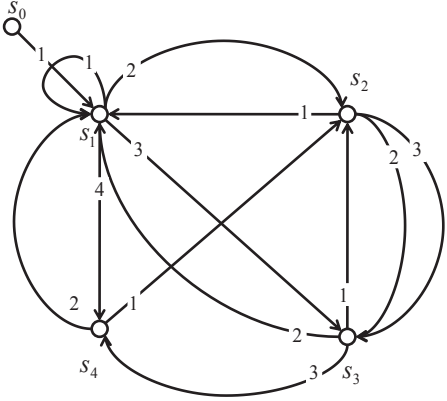}
\caption{The sequence graph $G$ with new labels.}
\label{fig_orderTransform2}
\end{figure}

By permutating the labels of the outgoing edges of $s_1$, we can have the graph as shown in Fig. \ref{fig_orderTransform2}.  The new labeling on $G$ is also a complete walk ending at state
$s_1$, and its path is $$s_0s_1s_1s_2s_1s_3s_2s_3s_1s_4s_2s_3s_4s_1$$

Based on Lemma \ref{lemma_graphTransform}, we have the following result
\begin{Lemma}Given a starting state $s_\alpha$ and two collections of sequences $\Lambda=[\Lambda_1,\Lambda_2,...,\Lambda_k,...,\Lambda_n]$ and $\Gamma=[\Lambda_1,...,\Gamma_k,...,\Lambda_n]$
such that $\Gamma_k\tmu \Lambda_k$ (tail-fixed permutation). Then $(s_\alpha, \Lambda)$ and $(s_\alpha, \Gamma)$ have the same feasibility. \label{lemma_graphTransform2}
\end{Lemma}

\proof We prove that if $(s_\alpha, \Lambda)$ is feasible, then $(s_\alpha, \Gamma)$ is also feasible. If $(s_\alpha, \Lambda)$ is feasible,  there exists a sequence $X$ such that $s_\alpha=x_1$ and $\Lambda=\pi(X)$. Suppose its last element is $x_N=s_\chi$.

When $k=\chi$, according to Lemma \ref{lemma_graphTransform}, we know that $(s_\alpha, \Gamma)$ is feasible.

When $k\neq \chi$, we assume that $\Lambda_k=\pi_k(X)=x_{k_1}x_{k_2}...x_{k_w}$. Let's consider the subsequence $\overline{X}=x_1x_2...x_{k_w-1}$ of $X$. Then $\pi_k(\overline{X})=\Lambda_k^{|\Lambda_k|-1}$ and the last element of $\overline{X}$ is $s_k$. According to Lemma \ref{lemma_graphTransform}, we can get that:
there exists a sequence  $x'_1x_2'...x_{k_w-1}'$ with $x'_1=x_1$ and $x_{k_w-1}'=x_{k_w-1}$ such that
$$\pi(x'_1x_2'...x_{k_w-1}')=[\pi_1(\overline{X}),...,\Gamma_k^{|\Gamma_k|-1},\pi_{k+1}(\overline{X}),...,\pi_n(\overline{X})]$$
since
$\Gamma_k^{|\Gamma_k|-1}\pmu \Lambda_k^{|\Lambda_k|-1}$.

Let $x_{k_w}'x_{k_w+1}'...x_{N}'=x_{k_w}x_{k_w+1}...x_{N}$, i.e., concatenating $x_{k_w}x_{k_w+1}...x_{N}$ to the end of $x_1'x_2'...x_{k_w-1}'$, we
can generate a sequence $x_1'x_2'...x_N'$ such that its exit sequence of state $s_k$ is
$$\Gamma_k^{|\Gamma_k|-1}*x_{k_w}=\Gamma_k$$
and its exit sequence of state $s_i$ with $i\neq k$ is $\Lambda_i=\pi_i(X)$.

So if $(s_\alpha, \Lambda)$ is feasible, then $(s_\alpha, \Gamma)$ is also feasible. Similarly, if $(s_\alpha, \Gamma)$ is
feasible, then $(s_\alpha, \Lambda)$ is feasible. As a result, $(s_\alpha, \Lambda)$ and $(s_\alpha, \Gamma)$ have the same feasibility.
\hfill\QED

According to the lemma above, we know that
$(s_\alpha, [\Lambda_1,\Lambda_2,...,\Lambda_n])$ and $(s_\alpha, [\Gamma_1,\Lambda_2,...,\Lambda_n])$ have the same feasibility,
$(s_\alpha, [\Gamma_1,\Lambda_2,...,\Lambda_n])$ and $(s_\alpha, [\Gamma_1,\Gamma_2,...,\Lambda_n])$ have the same feasibility, ...,
$(s_\alpha, [\Gamma_1,\Gamma_2,...,\Gamma_{n-1}, \Lambda_n])$ and $(s_\alpha, [\Gamma_1,\Gamma_2,...,\Gamma_{n-1},\Gamma_n])$ have the same feasibility, so
the statement in the main lemma is true.

\ifCLASSOPTIONcaptionsoff
  \newpage
\fi

\end{document}